\documentclass[aps,preprint]{revtex4}%
\usepackage{amsfonts}
\usepackage{amsmath}
\usepackage{amssymb}
\usepackage{graphicx}%
\begin{document}
\preprint{ }
\title[ ]{New Class of Solvable and Integrable Many-Body Models}
\author{R. W. Richardson}
\affiliation{Physics Deepartment, New York University, New York, NY, 10003}
\author{}
\affiliation{}
\author{}
\affiliation{}
\keywords{}
\pacs{}
\pacs{}
\pacs{}
\pacs{}
\pacs{}
\pacs{}
\pacs{}
\pacs{}
\pacs{}
\pacs{}
\pacs{}
\pacs{}
\pacs{PACS number}
\pacs{PACS number}
\pacs{}
\pacs{}
\pacs{}
\pacs{}
\pacs{}
\pacs{}
\pacs{}
\pacs{}
\pacs{}
\pacs{}
\pacs{}
\pacs{}
\pacs{}
\pacs{}
\pacs{}
\pacs{}
\pacs{}
\pacs{}
\pacs{PACS number}
\pacs{PACS number}
\pacs{}
\pacs{}
\pacs{}
\pacs{}
\pacs{}
\pacs{}
\pacs{}
\pacs{}
\pacs{}
\pacs{}
\pacs{}
\pacs{}
\pacs{}
\pacs{}
\pacs{}
\pacs{}
\pacs{}
\pacs{}
\pacs{PACS number}
\pacs{PACS number}
\pacs{}
\pacs{}
\pacs{}
\pacs{}
\pacs{}
\pacs{}
\pacs{}
\pacs{}
\pacs{}
\pacs{}
\pacs{}
\pacs{}
\pacs{}
\pacs{}
\pacs{}
\pacs{}
\pacs{}
\pacs{}
\pacs{PACS number}
\pacs{PACS number}
\pacs{}
\pacs{}
\pacs{}
\pacs{}
\pacs{}
\pacs{}
\pacs{}
\pacs{}
\pacs{}
\pacs{}
\pacs{}
\pacs{}
\pacs{}
\pacs{}
\pacs{}
\pacs{}
\pacs{}
\pacs{}
\pacs{PACS number}
\pacs{PACS number}
\pacs{}
\pacs{}
\pacs{}
\pacs{}
\pacs{}
\pacs{}
\pacs{}
\pacs{}
\pacs{}
\pacs{}
\pacs{}
\pacs{}
\pacs{}
\pacs{}
\pacs{}
\pacs{}
\pacs{}
\pacs{}
\pacs{PACS number}
\pacs{PACS number}
\pacs{}
\pacs{}
\pacs{}
\pacs{}
\pacs{}
\pacs{}
\pacs{}
\pacs{}
\pacs{}
\pacs{}
\pacs{}
\pacs{}
\pacs{}
\pacs{}
\pacs{}
\pacs{}
\pacs{}
\pacs{}
\pacs{PACS number}
\pacs{PACS number}
\pacs{}
\pacs{}
\pacs{}
\pacs{}
\pacs{}
\pacs{}
\pacs{}
\pacs{}
\pacs{}
\pacs{}
\pacs{}
\pacs{}
\pacs{}
\pacs{}
\pacs{}
\pacs{}
\pacs{}
\pacs{}
\pacs{PACS number}
\pacs{PACS number}
\pacs{}
\pacs{}
\pacs{}
\pacs{}
\pacs{}
\pacs{}
\pacs{}
\pacs{}
\pacs{}
\pacs{}
\pacs{}
\pacs{}
\pacs{}
\pacs{}
\pacs{}
\pacs{}
\pacs{}
\pacs{}
\pacs{PACS number}
\pacs{PACS number}
\pacs{}
\pacs{}
\pacs{}
\pacs{}
\pacs{}
\pacs{}
\pacs{}
\pacs{}
\pacs{}
\pacs{}
\pacs{}
\pacs{}
\pacs{}
\pacs{}
\pacs{}
\pacs{}
\pacs{}
\pacs{}
\pacs{PACS number}
\pacs{PACS number}
\pacs{}
\pacs{}
\pacs{}
\pacs{}
\pacs{}
\pacs{}
\pacs{}
\pacs{}
\pacs{}
\pacs{}
\pacs{}
\pacs{}
\pacs{}
\pacs{}
\pacs{}
\pacs{}
\pacs{}
\pacs{}
\pacs{PACS number}
\pacs{PACS number}
\pacs{}
\pacs{}
\pacs{}
\pacs{}
\pacs{}
\pacs{}
\pacs{}
\pacs{}
\pacs{}
\pacs{}
\pacs{}
\pacs{}
\pacs{}
\pacs{}
\pacs{}
\pacs{}
\pacs{}
\pacs{}
\pacs{PACS number}
\pacs{PACS number}
\pacs{}
\pacs{}
\pacs{}
\pacs{}
\pacs{}
\pacs{}
\pacs{}
\pacs{}
\pacs{}
\pacs{}
\pacs{}
\pacs{}
\pacs{}
\pacs{}
\pacs{}
\pacs{}
\pacs{}
\pacs{}
\pacs{PACS number}
\pacs{PACS number}
\pacs{}
\pacs{}
\pacs{}
\pacs{}
\pacs{}
\pacs{}
\pacs{}
\pacs{}
\pacs{}
\pacs{}
\pacs{}
\pacs{}
\pacs{}
\pacs{}
\pacs{}
\pacs{}
\pacs{}
\pacs{}
\pacs{PACS number}
\pacs{PACS number}
\pacs{}
\pacs{}
\pacs{}
\pacs{}
\pacs{}
\pacs{}
\pacs{}
\pacs{}
\pacs{}
\pacs{}
\pacs{}
\pacs{}
\pacs{}
\pacs{}
\pacs{}
\pacs{}
\pacs{}
\pacs{}
\pacs{PACS number}
\pacs{PACS number}
\pacs{}
\pacs{}
\pacs{}
\pacs{}
\pacs{}
\pacs{}
\pacs{}
\pacs{}
\pacs{}
\pacs{}
\pacs{}
\pacs{}
\pacs{}
\pacs{}
\pacs{}
\pacs{}
\pacs{}
\pacs{}
\pacs{PACS number}
\pacs{PACS number}
\pacs{}
\pacs{}
\pacs{}
\pacs{}
\pacs{}
\pacs{}
\pacs{}
\pacs{}
\pacs{}
\pacs{}
\pacs{}
\pacs{}
\pacs{}
\pacs{}
\pacs{}
\pacs{}
\pacs{}
\pacs{}
\pacs{PACS number}
\pacs{PACS number}
\pacs{}
\pacs{}
\pacs{}
\pacs{}
\pacs{}
\pacs{}
\pacs{}
\pacs{}
\pacs{}
\pacs{}
\pacs{}
\pacs{}
\pacs{}
\pacs{}
\pacs{}
\pacs{}
\pacs{}
\pacs{}
\pacs{PACS number}
\pacs{PACS number}
\pacs{}
\pacs{}
\pacs{}
\pacs{}
\pacs{}
\pacs{}
\pacs{}
\pacs{}
\pacs{}
\pacs{}
\pacs{}
\pacs{}
\pacs{}
\pacs{}
\pacs{}
\pacs{}
\pacs{}
\pacs{}
\pacs{PACS number}
\pacs{PACS number}
\pacs{}
\pacs{}
\pacs{}
\pacs{}
\pacs{}
\pacs{}
\pacs{}
\pacs{}
\pacs{}
\pacs{}
\pacs{}
\pacs{}
\pacs{}
\pacs{}
\pacs{}
\pacs{}
\pacs{}
\pacs{}
\pacs{PACS number}
\pacs{PACS number}
\pacs{}
\pacs{}
\pacs{}
\pacs{}
\pacs{}
\pacs{}

\begin{abstract}

\end{abstract}
\volumeyear{ }
\volumenumber{ }
\issuenumber{ }
\eid{ }
\date{}
\received[Received text]{}

\revised[Revised text]{}

\accepted[Accepted text]{}

\published[Published text]{}

\startpage{1}

\begin{center}
\textbf{New Class of Solvable and Integrable Many-Body Models}

R. W. Richardson

\emph{Physics Department, New York University, New York, NY, 10003}

\emph{e-mail: rwr1@nyu.edu}

\bigskip
\end{center}

PACS: 02.30.Ik, 67.40.Ik, 71.10.Li, 74.20.Fg

\bigskip

\begin{center}
\textbf{Abstract}
\end{center}

Integrability conditions for systems of bosons or fermions with seniority
conserving hamiltonians are derived. The conditions are shown to be invariant
under a large class of transformations of the interaction matrix elements.
Previously published integrable models are shown to satisfy these conditions
and the existence of a new class of integrable models is demonstrated. The
number of free parameters in the interaction in these models equals the number
of single particle levels plus 3 . Equations for the energy eigenstates and
eigenvalues are derived and the eigenvalues of the complete set of two-body
integrals of the motion are given for the new class. Some two-body
correlations in these eigenstates are derived from the integrals of the motion.

\bigskip

\section{Introduction}

\qquad Shortly after the development of the BCS\ theory of
superconductivity\cite{1}, the Hamiltonian and the associated computational
techniques were adapted to create the pairing model of nuclear structure. This
model has been successful in qualitatively accounting for nuclear properties
such as the energy gap in the single-particle spectrum of even-even nuclei,
odd-even mass differences, and the moments of inertia of deformed nuclei in
spite of the fact that the methods used by BCS do not apply to finite systems
since they violate number conservation. A great deal of effort was invested
into attempts to restore number conservation to the theory. At about the same
time, an exact analytical expression\cite{2,3} for the eigenstates and
eigenvalues of the pairing Hamiltonian with constant interaction matrix
elements was published and largely ignored by the nuclear physics community.
Shortly thereafter the corresponding many-boson problem was also
solved\cite{4}. More recently, the attention of the condensed matter community
has been directed towards understanding pairing correlations in finite systems
by a series of experiments on metallic grains with a size of a few
nanometers\cite{5,6} and the exact solution has been extensively utilized in
studies of these systems\cite{6}. A separate and important recent development
has been the demonstration that the pairing Hamiltonian, with constant
interaction matrix elements, is an integrable system.\cite{7} This system is
then both solvable and integrable. Another important result is the
demonstration of a class of solvable and integrable models involving both
pairing forces and an interaction term quadratic in the particle number
operators\cite{8,9}. This paper draws much of its inspiration from these works
and extends their results.

\qquad The structure of this paper is as follows. We start in Sec. 2 with a
definition of the class of Hamiltonians that we will consider. They are the
most general, seniority-conserving Hamiltonians with two-body forces, which
describe systems of fermions or bosons. These Hamiltonians have a diagonal
single-particle kinetic energy plus a pairing term and a term quadratic in the
particle number operators with arbitrary matrix elements in all terms. We then
consider a set of operators that have the same structure as the Hamiltonian
and derive a set of conditions on the matrix elements in the members of the
set that must be satisfied if these operators are required to commute with the
Hamiltonian. We obtain an explicit solution for the matrix elements in the
two-body part of these constants of the motion in terms of the matrix elements
in the Hamiltonian and a set of conditions which the Hamiltonian matrix
elements must satisfy. Solutions of these conditions yield integrable
Hamiltonians. We show that, due to a symmetry of the conditions, each
particular solution gives rise to a large family of solutions. The number of
free parameters in each family is greater than the number of single-particle
energy levels. We also show that previously published\cite{8,9} solvable and
integrable models satisfy our conditions. By generating a new particular
solution of the conditions, we generate a whole new family of integrable
models. In Sec. 3, we show that members of this new family of integrable
models are also solvable by giving analytical forms for all their eigenstates
and eigenvalues. In Sec. 4, we calculate the occupations of the
single-particle levels and some two-body correlation functions in these eigenstates.

\bigskip

\section{Hamiltonian and Constants of the Motion}

\qquad We consider the class of seniority conserving Hamiltonians for Fermi or
Bose systems that can be written in the following form $H=T+U+V$ , where%

\begin{align}
T  &  =\sum_{i}2\epsilon_{i}~n_{i}\tag{2.1}\\
U  &  =-\frac{sg}{4}\sum_{i,j}u_{i,j}\left(  \Omega_{i}+2~s~n_{i}\right)
\left(  \Omega_{j}+2~s~n_{j}\right) \nonumber\\
V  &  =\frac{g}{2}\sum_{i,j}v_{i,j}\left(  b_{i}^{\dag}~b_{j}+b_{j}^{\dag
}~b_{i}\right) \nonumber
\end{align}

In these expressions, $i$ labels a single-particle energy-level of energy
$\epsilon_{i}$ and $g$ is an interaction strength. The interaction matrices
$u$ and $v$ are real and symmetric. The statistics factor $s=-1\left(
+1\right)  $ for Fermi (Bose) statistics. The operators $n$ and $b$ are
defined by

\qquad Fermions:

The degeneracy of level $i$ is $2\Omega_{i}$ with the factor $2$ coming from
time-reversal degeneracy and $\alpha=\pm1,\cdot\cdot\cdot,\pm\Omega_{i}$ with%

\begin{align}
n_{i}  &  =\frac{1}{2}\sum_{\alpha=1}^{\Omega_{i}}\left(  a_{i,\alpha}^{\dag
}a_{i,\alpha}+a_{i,-\alpha}^{\dag}a_{i,-\alpha}\right) \tag{2.2}\\
b_{i}  &  =\sum_{\alpha=1}^{\Omega_{i}}a_{i,-\alpha}a_{i,\alpha}\nonumber
\end{align}

\qquad Bosons:

There are 2 possibilities for bosons. In the first case, the single-particle
states are not time-reversal eigenstates and the above definition applies. In
the second case, the single-particle states are time-reversal eigenstates as
in the $\mathbf{k=0}$ state or the bound states of a potential. We then define
the operators%

\begin{align}
n_{i}  &  =\frac{1}{2}\sum_{\alpha=1}^{\Omega_{i}}a_{i,\alpha}^{\dag
}a_{i,\alpha}\tag{2.3}\\
b_{i}  &  =\frac{1}{\sqrt{2}}\sum_{\alpha=1}^{\Omega_{i}}a_{i,\alpha
}a_{i,\alpha}\nonumber
\end{align}

\qquad The single-particle operators $a$ satisfy (Fermi)Bose (anti)commutation
relations which can be used to derive the commutation rules for the $n$ and
$b$ operators%

\begin{align}
\left[  n_{i}\,,\,b_{j}^{\dag}\right]   &  =\delta_{i,j}b_{j}^{\dag}%
\tag{2.4}\\
\left[  b_{i}\,,\,b_{j}^{\dag}\right]   &  =\delta_{i,j}\left(  \Omega
_{j}+2s~n_{j}\right) \nonumber
\end{align}

The effects of degeneracy are expressed explicitly in the presence of the
$\Omega_{j}$ in the commutation relations and the effects of statistics in the
presence of the $s$ in $U$ and in the factors $\left(  \Omega_{j}%
+2s~n_{j}\right)  $ in the Hamiltonian and the commutation relations. The
seniority operators are defined by%

\[
\nu_{i}=\sum_{\alpha=1}^{\Omega_{i}}\left(  a_{i,\alpha}^{\dag}a_{i,\alpha
}-a_{i,-\alpha}^{\dag}a_{i,-\alpha}\right)
\]

and they commute with the Hamiltonian. They count the number of unpaired
particles in each single particle level. The form of the interaction in (2.1)
is chosen so as to mimic that of a Heisenberg XXZ spin chain model after a
quasispin substitution for the fermion pair operators.

\qquad The eigenstates of such Hamiltonians factor into a product of a state
of the unpaired particles times a state of the paired particles. The energy of
the state is the sum of the energies of the unpaired particles, which interact
only through $U$, plus that of the paired particles. The influence of the
unpaired particles on the paired particles comes about through through a
modification of the degeneracies $\Omega_{i}$ - blocking in the case of
fermions and antiblocking for bosons. In what follows, we only consider the
seniority-zero paired particles and assume that the degeneracies are effective
degeneracies which depend upon which state of the system is being considered.

\qquad Following Ref. 7, we look for two-body constants of the motion $K$ that
have a form similar to that of the Hamiltonian, $K=X+Y+Z$ , with%

\begin{align}
X  &  =\sum_{j}2x_{j}~n_{j}\tag{2.5}\\
Y  &  =-\frac{sg}{4}\sum_{j,k}y_{j,k}\left(  \Omega_{j}+2~s~n_{j}\right)
\left(  \Omega_{k}+2~s~n_{k}\right) \nonumber\\
Z  &  =\frac{g}{2}\sum_{j,k}z_{j,k}\left(  b_{j}^{\dag}~b_{k}+b_{k}^{\dag
}~b_{j}\right) \nonumber
\end{align}

where $y$ and $z$ are real and symmetric matrices to be determined in terms of
$x$ , $\epsilon$ , $u$ , and $v$ so that $K$ commutes with the Hamiltonian.
Since $x$ is a set of free parameters, equal in number to the number of
degrees of freedom of the system, the solution of the resulting equations will
provide a complete set of constants of the motion specified by a choice of $x$
and it will be shown that these constants, with different choices for $x$ ,
commute with each other.

\qquad The calculation of the commutator $\left[  H\,,\,K\right]  $ follows
from (2.1), (2.4), and (2.5) and some tedious algebra. The result is%

\begin{align}
&  \left[  H~,~K\right] \tag{2.6}\\
&  =g\sum_{k,l}\left\{  -\left(  \xi_{k}-\xi_{l}\right)  v_{k,l}+\left(
e_{k}-e_{l}\right)  z_{k,l}\right\}  ~\left(  b_{k}^{\dag}~b_{l}-b_{l}^{\dag
}~b_{k}\right) \nonumber\\
&  +\frac{g^{2}}{2}\sum_{j,k,l}\left[  v_{k,l}\left(  y_{j,k}-y_{j,l}\right)
-\left(  u_{j,k}-u_{j,l}\right)  z_{k,l}+\left(  v_{j,k}z_{j,l}-v_{j,l}%
z_{j,k}\right)  \right]  \left[  b_{k}^{\dag}~\left(  \Omega_{j}%
+2~s~n_{j}\right)  ~b_{l}-b_{l}^{\dag}~\left(  \Omega_{j}+2~s~n_{j}\right)
~b_{k}\right] \nonumber
\end{align}

where%

\begin{align}
\xi_{k}  &  =x_{k}-%
\frac12
\ s\ g\ y_{k,k}\tag{2.7}\\
e_{k}  &  =\epsilon_{k}-%
\frac12
\ s\ g\ u_{k,k}\nonumber
\end{align}

Setting the commutator equal to zero yields the equations%

\begin{equation}
z_{k,l}=\frac{\left(  \xi_{k}-\xi_{l}\right)  }{\left(  e_{k}-e_{l}\right)
}v_{k,l}~,~k\neq l \tag{2.8}%
\end{equation}

and%

\begin{equation}
v_{k,l}\left(  y_{j,k}-y_{j,l}\right)  -\left(  u_{j,k}-u_{j,l}\right)
z_{k,l}+\left(  v_{j,k}z_{j,l}-v_{j,l}z_{j,k}\right)  =0~,~k\neq l \tag{2.9}%
\end{equation}

\qquad We assume that $v\neq0$ and first consider the case in which $u=0$ .
Then (2.9), with $j\rightarrow k$ , is%

\begin{equation}
v_{k,l}\left(  y_{k,k}-z_{k,k}-y_{k,l}\right)  +v_{k,k}z_{k,l}=0~,~k\neq l
\tag{2.10}%
\end{equation}

Without loss of generality, we can require%

\begin{equation}
y_{k,k}=z_{k,k}=0 \tag{2.11}%
\end{equation}

Then%

\begin{equation}
y_{k,l}=\frac{\left(  x_{k}-x_{l}\right)  }{\left(  e_{k}-e_{l}\right)
}v_{k,l}~,~k\neq l \tag{2.12}%
\end{equation}

which requires that $v_{k,k}$ be a constant which can be taken to be one.
Equation (2.9), with $j$ , $k$ and $l$ distinct, and, using (2.8) and (2.12), becomes%

\[
\left(  v_{k,l}-v_{j,k}v_{j,l}\right)  \left(  \frac{\left(  x_{j}%
-x_{k}\right)  }{\left(  e_{j}-e_{k}\right)  }-\frac{\left(  x_{j}%
-x_{l}\right)  }{\left(  e_{j}-e_{l}\right)  }\right)  =0~,~j\neq k\neq l\neq
j
\]

which requires $v_{j,k}=1$ and is the simple pairing Hamiltonian with constant
interaction matrix elements. This is the result of Ref. 7.

\qquad We now assume that both $u\neq0$ and $v\neq0$ . We can solve (2.9),
which is antisymmetric in $k$ and $l$ , for $y$ by setting $j=k$ which gives%

\begin{equation}
v_{k,l}\left(  y_{k,k}-z_{k,k}\right)  -v_{k,l}y_{k,l}+\left(  u_{k,l}%
+v_{k,k}-u_{k,k}\right)  z_{k,l}=0~,~k\neq l \tag{2.13}%
\end{equation}

Without loss of generality, we can require%

\begin{equation}
y_{k,k}=z_{k,k}=v_{k,k}-u_{k,k}=0 \tag{2.14}%
\end{equation}

Then, using (2.8), we have%

\begin{equation}
y_{k,l}=\frac{\left(  x_{k}-x_{l}\right)  }{\left(  e_{k}-e_{l}\right)
}u_{k,l}~,~k\neq l \tag{2.15}%
\end{equation}

\qquad Equation (2.9), with $j$ , $k$ and $l$ distinct, and using (2.8) and
(2.15) provides an integrability condition on $u$ and $v$ . This condition is
linear in the $x^{\prime}s$ and, equating their coefficients to zero, leads to
the single equation%

\begin{equation}
\left(  x-y\right)  u\left(  x,z\right)  v\left(  y,z\right)  -\left(
x-z\right)  u\left(  x,y\right)  v\left(  y,z\right)  +\left(  y-z\right)
v\left(  x,y\right)  v\left(  x,z\right)  =0\ ,\ x\neq y\neq z\neq x
\tag{2.16}%
\end{equation}

where we have introduced the notation%

\begin{align}
u_{j,k}  &  =u\left(  e_{j},e_{k}\right) \tag{2.17}\\
v_{j,k}  &  =v\left(  e_{j},e_{k}\right) \nonumber
\end{align}

and write $\left(  x,y,z\right)  $ for $\left(  e_{j},e_{k},e_{l}\right)  $ .
Note that this equation is independent of particle statistics and
single-particle degeneracies.

\qquad Eq. (2.16) has a curious symmetry that allows the generation of a whole
class of solutions from any particular solution. Assume that we have a
particular solution of (2.16), $u\left(  x,y\right)  $ and $v\left(
x,y\right)  $ , then there will exist a class of solutions generated from this
particular solution of the form%

\begin{align}
u^{\prime}\left(  x,y\right)   &  =\frac{\left(  x-y\right)  }{f\left(
x\right)  -f\left(  y\right)  }u\left[  f\left(  x\right)  ,f\left(  y\right)
\right] \tag{2.18}\\
v^{\prime}\left(  x,y\right)   &  =\frac{\left(  x-y\right)  }{f\left(
x\right)  -f\left(  y\right)  }v\left[  f\left(  x\right)  ,f\left(  y\right)
\right] \nonumber
\end{align}

where $f$ is some well behaved function of its argument. This result follows
from the fact that the form of Eq. (2.16) is invariant under the transformation%

\begin{equation}
u\rightarrow u^{\prime}~,~v\rightarrow v^{\prime}~,~x\rightarrow f \tag{2.19}%
\end{equation}

Note that iteration of the transformation (2.18) is equivalent to a
redefinition of the function $f$ so that each class of solutions is closed
under this operation. If we require the functions $f$ to be monotonic with
single-valued inverses, then the transformations have a group structure. In
the following we will look for particular solutions of (2.16) which will
provide the basis for classes of solutions generated by the transformations
(2.18). The members of each class are specified by the set of numbers
$f_{i}=f\left(  e_{i}\right)  $ so that the number of free parameters in a
class is equal to the number of single-particle energy levels plus the number
of free parameters in the particular solution.

\qquad We first show that recently published\cite{8,9} results are solutions
of (2.16). \ The first class of solutions is based upon the simplest solution
of (2.16): $u=v=1$ . With this solution, $U$ is a function of the total number
of particles, and the Hamiltonian is equivalent to the simple pairing model
Hamiltonian with constant interaction matrix elements. The members of this
class, after the transformation (2.18), are the descendents of this
Hamiltonian with%

\begin{equation}
u\left(  x,y\right)  =v\left(  x,y\right)  =\frac{x-y}{f\left(  x\right)
-f\left(  y\right)  } \tag{2.20}%
\end{equation}

This class can also be obtained directly from (2.16) by taking $u=v$ and
dividing (2.16) by $v\left(  x,y\right)  v\left(  x,z\right)  v\left(
y,z\right)  $ to get the equation%

\begin{equation}
\frac{\left(  x-y\right)  }{v\left(  x,y\right)  }-\frac{\left(  x-z\right)
}{v\left(  x,z\right)  }+\frac{\left(  y-z\right)  }{v\left(  y,z\right)  }=0
\tag{2.21}%
\end{equation}

which is linear in $1/v$ . The solutions of this equation are $\left(
x-y\right)  /v\left(  x,y\right)  =f\left(  x\right)  -f\left(  y\right)  $
which is (2.20). A second class of solutions is obtained by writing%

\[
u\left(  x,y\right)  =v\left(  x,y\right)  w\left(  x,y\right)
\]

and obtaining from (2.16)%

\begin{equation}
\frac{\left(  x-y\right)  }{v\left(  x,y\right)  }w\left(  x,z\right)
-\frac{\left(  x-z\right)  }{v\left(  x,z\right)  }w\left(  x,y\right)
+\frac{\left(  y-z\right)  }{v\left(  y,z\right)  }=0 \tag{2.22}%
\end{equation}

Solutions of this equation are\cite{8,9}%

\begin{align}
v\left(  x,y\right)   &  =\frac{q\left(  x-y\right)  }{Sinh\left[  q\left(
x-y\right)  \right]  }\tag{2.23}\\
w\left(  x,y\right)   &  =Cosh\left[  q\left(  x-y\right)  \right] \nonumber\\
u\left(  x,y\right)   &  =\frac{q\left(  x-y\right)  Cosh\left[  q\left(
x-y\right)  \right]  }{Sinh\left[  q\left(  x-y\right)  \right]  }\nonumber
\end{align}

Note that the first class of solutions is just the $q\rightarrow0$ limit of
the second.

\qquad A third and new class of solutions is obtained by writing%

\begin{equation}
v\left(  x,y\right)  =\gamma\left(  x\right)  \gamma\left(  y\right)
\tag{2.24}%
\end{equation}

where $\gamma^{2}\left(  x\right)  $ is a polynomial in $x$ and $u\left(
x,y\right)  $ is a symmetric polynomial in $x$ and $y$ . Substituting this
form into (2.16) gives the equation%

\begin{equation}
\left(  x-y\right)  u\left(  x,z\right)  -\left(  x-z\right)  u\left(
x,y\right)  +\left(  y-z\right)  \gamma^{2}\left(  x\right)  =0 \tag{2.25}%
\end{equation}

which is linear in $u$ and $\gamma^{2}$ . The solution of this equation is%

\begin{align}
u\left(  x,y\right)   &  =1+\gamma_{1}\ \left(  x+y\right)  +\gamma
_{2}\ x\ y\tag{2.26}\\
\gamma\left(  x\right)   &  =\left(  1+2\ \gamma_{1}\ x+\gamma_{2}%
\ x^{2}\right)  ^{%
\frac12
}\nonumber
\end{align}

with $\gamma_{1}$ and $\gamma_{2}$ free parameters. Note that before the
transformation (2.19), the first two terms in $u$ can be ignored since they
result in terms in the Hamiltonian that are either constants, proportional to
the total number of particles, or a modification of the single-particle
spectrum. However, they are important after the transformation. Note also that
this solution reverts back to the first one for the case $\gamma_{2}%
=\gamma_{1}^{2}$ . The case in which there is no constant term in $u$ can be
handled by making the replacements $g\rightarrow g\gamma_{0}$ , $\gamma
_{i}\rightarrow\gamma_{i}/\gamma_{0}$ , $i=1,2$ , and then taking the limit
$\gamma_{0}\rightarrow0$ . The net effect of which is the replacement of $1$
in (2.26) by $0$ .

\qquad Further study of Eq. (2.16) has not produced any other simple solutions
that are not the result of symmetry transformations applied to the solutions
(2.20), (2.23), and (2.26). In the next sections, we consider the eigenstates
of the integrable model (2.26) after a symmetry transformation.

\bigskip

\section{Eigenstates}

\qquad It is a semiempirical theorem that the seniority-zero eigenstates of
integrable Hamiltonians of the form (2.1), have the form of a product of
correlated pairs. This is true for the states of the Hamiltonians with
interaction matrix elements given by (2.20) or (2.23)\cite{8,9} and we now
show, by construction, that it is also true for the solution (2.26).

\qquad Assuming that the eigenstates can be taken as a product of correlated
pairs, we write, for a typical eigenstate of $2N$ particles,%

\begin{equation}
\left\vert \psi\right\rangle =\prod_{r=1}^{N}B_{r}^{\dag}\left\vert
0\right\rangle \tag{3.1}%
\end{equation}

where%

\begin{equation}
B_{r}^{\dag}=\sum_{j}\phi_{r}\left(  e_{j}\right)  b_{j}^{\dag} \tag{3.2}%
\end{equation}

Then, commuting the Hamiltonian past the product of $B^{\dag}$ 's , we have%

\begin{align}
\left(  H-2\sum_{p=1}^{N}E_{p}-E^{\left(  0\right)  }\right)  \left\vert
\psi\right\rangle  &  =\sum_{p=1}^{N}\left(  \prod_{r=1,\left(  r\neq
p\right)  }^{N}B_{r}^{\dag}\right)  \left\{  \left[  H\,,\,B_{p}^{\dag
}\right]  -2E_{p}B_{p}^{\dag}\right\}  \left\vert 0\right\rangle \tag{3.3}\\
&  +\frac{1}{2}\sum_{p,q=1\left(  p\neq q\right)  }^{N^{\prime}}\left(
\prod_{r=1,\left(  r\neq p,q\right)  }^{N}B_{r}^{\dag}\right)  \left[  \left[
H\,,\,B_{p}^{\dag}\right]  \,,\,B_{q}^{\dag}\right]  \left\vert 0\right\rangle
\nonumber
\end{align}

where we have written the energy of the vacuum state as%

\begin{equation}
E^{\left(  0\right)  }=-\frac{sg}{4}\sum_{i,j}u_{i,j}\ \Omega_{i}\ \Omega_{j}
\tag{3.4}%
\end{equation}

and where the parameters $2E_{p}$ are pair-energies to be determined. The
energy of the state will be given in terms of these parameters. The
commutators can be calculated from (2.4) and (3.2) with the results%

\begin{equation}
\left\{  \left[  H\,,\,B_{p}^{\dag}\right]  -2E_{p}B_{p}^{\dag}\right\}
\left\vert 0\right\rangle =2\sum_{j}\left[  \left(  e_{j}-\frac{g}{2}\sum
_{k}u_{j,k}\ \Omega_{k}-E_{p}\right)  \,\phi_{p}\left(  e_{j}\right)
+\frac{g}{2}\sum_{k}v_{j,k}\ \Omega_{k}\ \phi_{p}\left(  e_{k}\right)
\right]  \,b_{j}^{\dag}\left\vert 0\right\rangle \tag{3.5}%
\end{equation}

\begin{equation}
\left[  \left[  H\,,\,B_{p}^{\dag}\right]  \,,\,B_{q}^{\dag}\right]
=-gs\sum_{j,k}\left\{  u_{j,k}\left[  \phi_{p}\left(  e_{j}\right)  \phi
_{q}\left(  e_{k}\right)  +\phi_{p}\left(  e_{k}\right)  \phi_{q}\left(
e_{j}\right)  \right]  -v_{j,k}\left[  \phi_{p}\left(  e_{j}\right)  \phi
_{q}\left(  e_{j}\right)  +\phi_{p}\left(  e_{k}\right)  \phi_{q}\left(
e_{k}\right)  \right]  \right\}  b_{j}^{\dag}b_{k}^{\dag} \tag{3.6}%
\end{equation}

\qquad For the new set of integrable models, we take $u$ and $v$ to be given
by (2.26). Then, after the symmetry transformation (2.18),%

\begin{align}
u_{j,k}  &  =u\left(  e_{j},e_{k}\right)  \rightarrow\frac{\left(  e_{j}%
-e_{k}\right)  }{\left(  f_{j}-f_{k}\right)  }u\left(  f_{j},f_{k}\right)
\tag{3.7}\\
v_{j,k}  &  =v\left(  e_{j},e_{k}\right)  =\gamma\left(  e_{j}\right)
\gamma\left(  e_{k}\right)  \rightarrow\frac{\left(  e_{j}-e_{k}\right)
}{\left(  f_{j}-f_{k}\right)  }\gamma\left(  f_{j}\right)  \gamma\left(
f_{k}\right) \nonumber\\
\phi_{p}\left(  e_{k}\right)   &  =\frac{\gamma\left(  e_{k}\right)  }%
{e_{k}-\omega_{p}}\rightarrow\phi_{p}\left(  f_{k}\right)  =\frac
{\gamma\left(  f_{k}\right)  }{f_{k}-\omega_{p}}\nonumber
\end{align}

with $\omega_{p}$ a parameter to be determined. For (3.5), we then have%

\begin{align}
\left\{  \left[  H\,,\,B_{p}^{\dag}\right]  -2E_{p}B_{p}^{\dag}\right\}
\left\vert 0\right\rangle  &  =2\sum_{j}\left[  \left(  e_{j}-\frac{g}{2}%
\sum_{k}\ \Omega_{k}\frac{\left(  e_{j}-e_{k}\right)  }{\left(  f_{j}%
-f_{k}\right)  }\left[  u\left(  f_{j},f_{k}\right)  -\gamma^{2}\left(
f_{k}\right)  \right]  -E_{p}\right)  \right. \tag{3.8}\\
&  \left.  +\frac{g}{2}\sum_{k}\ \Omega_{k}\left(  e_{j}-e_{k}\right)
\frac{\gamma^{2}\left(  f_{k}\right)  }{\left(  f_{k}-\omega_{p}\right)  }%
\phi_{p}\left(  f_{j}\right)  \,b_{j}^{\dag}\right]  \left\vert 0\right\rangle
\nonumber\\
&  =2\sum_{j}\left[  e_{j}+\frac{g}{2}\sum_{k}\Omega_{k}\left(  e_{j}%
-e_{k}\right)  \frac{u\left(  \omega_{p},f_{k}\right)  }{\left(  f_{k}%
-\omega_{p}\right)  }-E_{p}\right]  \,\phi_{p}\left(  e_{j}\right)
\,b_{j}^{\dag}\left\vert 0\right\rangle \nonumber
\end{align}

and for (3.6), after algebraic simplification,%

\begin{align}
\left[  \left[  H\,,\,B_{p}^{\dag}\right]  \,,\,B_{q}^{\dag}\right]   &
=-gs\frac{u\left(  \omega_{p},\omega_{q}\right)  }{\left(  \omega_{p}%
-\omega_{q}\right)  }\sum_{j,k}\left(  e_{j}-e_{k}\right)  \left[  \phi
_{p}\left(  f_{j}\right)  \phi_{q}\left(  f_{k}\right)  -\phi_{p}\left(
f_{k}\right)  \phi_{q}\left(  f_{j}\right)  \right]  b_{j}^{\dag}b_{k}^{\dag
}\tag{3.9}\\
&  =-2gs\frac{u\left(  \omega_{p},\omega_{q}\right)  }{\left(  \omega
_{p}-\omega_{q}\right)  }\sum_{j}e_{j}\left[  \phi_{p}\left(  f_{j}\right)
B_{q}^{\dag}-\phi_{q}\left(  f_{j}\right)  B_{p}^{\dag}\right]  b_{j}^{\dag
}\nonumber
\end{align}

Substituting these results into (3.3) yields%

\begin{align}
&  \left(  H-2\sum_{p=1}^{N}E_{p}-E^{\left(  0\right)  }\right)  \left\vert
\psi\right\rangle \tag{3.10}\\
&  =2\sum_{p=1}^{N}\left(  \prod_{r=1,\left(  r\neq p\right)  }^{N}B_{r}%
^{\dag}\right)  \sum_{j}\left[  e_{j}\left(  1+\frac{g}{2}\sum_{k}\Omega
_{k}\frac{u\left(  \omega_{p},f_{k}\right)  }{\left(  f_{k}-\omega_{p}\right)
}-gs\sum_{q=1\left(  q\neq p\right)  }^{N^{\prime}}\frac{u\left(  \omega
_{p},\omega_{q}\right)  }{\left(  \omega_{p}-\omega_{q}\right)  }\right)
\right. \nonumber\\
&  \left.  -\frac{g}{2}\sum_{k}\Omega_{k}e_{k}\frac{u\left(  \omega_{p}%
,f_{k}\right)  }{\left(  f_{k}-\omega_{p}\right)  }-E_{p}\,\phi_{p}\left(
f_{j}\right)  \,b_{j}^{\dag}\right]  \left\vert 0\right\rangle \nonumber
\end{align}

Setting this equal to zero yields the equations%

\begin{equation}
\left[  1+\frac{g}{2}\sum_{k}\ \Omega_{k}\frac{u\left(  \omega_{p}%
,f_{k}\right)  }{\left(  f_{k}-\omega_{p}\right)  }+gs\sum_{q=1\left(  q\neq
p\right)  }^{N^{\prime}}\frac{u\left(  \omega_{p,}\omega_{q}\right)  }{\left(
\omega_{q}-\omega_{p}\right)  }\right]  =0~,~p=1,\cdot\cdot\cdot,N \tag{3.11}%
\end{equation}

\begin{equation}
E_{p}=-\frac{g}{2}\sum_{k}\ \Omega_{k}e_{k}\frac{u\left(  \omega_{p}%
,f_{k}\right)  }{\left(  f_{k}-\omega_{p}\right)  } \tag{3.12}%
\end{equation}

The energy of the state is then given by%

\begin{align}
E  &  =E^{\left(  0\right)  }+2\sum_{p=1}^{N}E_{p}\tag{3.13}\\
&  =E^{\left(  0\right)  }-g\sum_{p=1}^{N}\sum_{k}\ \Omega_{k}e_{k}%
\frac{u\left(  \omega_{p},f_{k}\right)  }{\left(  f_{k}-\omega_{p}\right)
}\nonumber
\end{align}

For $f_{k}=e_{k}$ , this expression simplifies to%

\begin{equation}
E=E^{\left(  0\right)  }+2\sum_{p=1}^{N}\omega_{p}-gs\sum_{p,q=1\left(  q\neq
p\right)  }^{N^{\prime}}u\left(  \omega_{p,}\omega_{q}\right)  -g\sum
_{p=1}^{N}\sum_{k}\ \Omega_{k}u\left(  \omega_{p},e_{k}\right)  \tag{3.14}%
\end{equation}

\qquad The equations (3.11) can be transformed into a form that allows a
two-dimensional electrostatic analogy for fermions with $s=-1$ Some algebraic
manipulation of (3.11) yields the equations%

\begin{equation}
-\frac{2+g\left[  \Omega+2s\left(  N-1\right)  \right]  \left(  \gamma
_{1}\ +\gamma_{2}\ \omega_{p}\right)  }{2g\gamma^{2}\left(  \omega_{p}\right)
}-\frac{1}{2}\sum_{k}\ \frac{\Omega_{k}}{\left(  f_{k}-\omega_{p}\right)
}-s\sum_{q=1\left(  q\neq p\right)  }^{N^{\prime}}\frac{1}{\left(  \omega
_{q}-\omega_{p}\right)  }=0~,~p=1,\cdot\cdot\cdot,N \tag{3.15}%
\end{equation}

For fermions, these equations can be interpreted as the equations for the
unstable equilibrium positions of $N$ unit line charges at the locations
$\omega_{p}$ in the complex plane interacting with charges $-\Omega_{k}/2$ at
the locations $f_{k}$ and two charges located at the zeros of $\gamma
^{2}\left(  \omega\right)  $ . In this analogy, the statistical repulsion
between fermions is mapped onto the electrostatic repulsion of like charges in
two dimensions. Such equations have been shown to lead to the BCS equations
plus corrections in a carefully defined $N\rightarrow\infty$ limit.\cite{10}
Eq. (3.14) then shows that the energy of the states has terms proportional to
the dipole and quadrupole moments of the free charge distribution. For bosons,
we have a statistical attraction between particles and the electrostatic
analogy does not work. Such equations have been shown to yield a generalized
Bose condensation into the lowest two single-particle levels for a repulsive
interaction in the $N\rightarrow\infty$ limit.\cite{4} The Bogoliubov
approximation, modified to accommodate this generalized condensation, then
gives good results in this limit.

\qquad The same techniques can be used to calculate the eigenvalues of the
integrals $K$ . We consider the solution (2.26) after a symmetry
transformation. We then have%

\begin{equation}
K\ \left\vert \psi\right\rangle =\left[  K^{\left(  0\right)  }-g\sum
_{p=1}^{N}\sum_{k}\Omega_{k}x_{k}\frac{u\left(  f_{k},\omega_{p}\right)
}{\left(  f_{k}-\omega_{p}\right)  }\right]  \left\vert \psi\right\rangle
\tag{3.16}%
\end{equation}

where%

\begin{equation}
K^{\left(  0\right)  }=-\frac{sg}{4}\sum_{j,k}\Omega_{j}\ \Omega_{k}\ y_{j,k}
\tag{3.17}%
\end{equation}

If we make the choice $x_{k}\rightarrow\delta_{i,k}$ and $K\rightarrow K_{i}$
, then $K_{i}$ and its eigenvalues, $k_{i}$ , are%

\begin{align}
K_{i}  &  =2\ n_{i}+g\sum_{k\left(  k\neq i\right)  }\frac{1}{\left(
f_{i}-f_{k}\right)  }\left[  v\left(  f_{i},f_{k}\right)  \left(  b_{i}^{\dag
}~b_{k}+b_{k}^{\dag}~b_{i}\right)  -\frac{s}{2}u\left(  f_{i},f_{k}\right)
\left(  \Omega_{i}+2~s~n_{i}\right)  \left(  \Omega_{k}+2~s~n_{k}\right)
\right] \tag{3.18}\\
k_{i}  &  =g\left[  K_{i}^{\left(  0\right)  }-\Omega_{i}\sum_{p=1}^{N}%
\frac{u\left(  f_{i},\omega_{p}\right)  }{\left(  f_{i}-\omega_{p}\right)
}\right] \nonumber\\
K_{i}^{\left(  0\right)  }  &  =-\frac{s}{2}\Omega_{i}\sum_{k\left(  k\neq
i\right)  }\Omega_{k}\frac{u\left(  f_{i},f_{k}\right)  }{\left(  f_{i}%
-f_{k}\right)  }\nonumber
\end{align}

This result will be used in the next section to calculate the occupations of
the single-particle levels and some two-body correlation functions in these eigenstates.

\bigskip

\section{Occupations and Correlations}

\qquad We can use the eigenvalues of the Hamiltonian and the $K$ ,s and the
Hellman-Feynman\cite{14} theorem to calculate certain two-body correlations in
the eigenstates presented in the previous section. The Hamiltonian and
constants of the motion are functions of the parameters in the potential, $g$
, $\gamma_{1}$ , and $\gamma_{2}$ as well as the single-particle spectrum
$e_{j}$ , and its transform $f_{j}=f\left(  e_{j}\right)  $ . Since the
function $f$ is unspecified, we can treat $e_{j}$ and $f_{j}$ as independent
parameters. In this picture the $\omega_{p}$ are functions of $g$ and the
$f_{j}$ , through (3.11), and do not depend upon the $e_{j}$ . Furthermore,
comparing (3.13) with (3.18) yields the result%

\begin{equation}
k_{i}-K^{\left(  0\right)  }=\frac{\partial}{\partial e_{i}}\left(
E-E^{\left(  0\right)  }\right)  \tag{4.1}%
\end{equation}

\qquad In order to calculate the mean occupations of the single-particle
levels, we use (3.18) to write%

\begin{equation}
n_{i}=\frac{1}{2}\left(  K_{i}-g\frac{\partial K_{i}}{\partial g}\right)
\tag{4.2}%
\end{equation}

The Hellman-Feynman theorem then yields%

\begin{align}
\left\langle \psi\right\vert n_{i}\left\vert \psi\right\rangle  &  =\frac
{1}{2}\left(  k_{i}-g\frac{\partial k_{i}}{\partial g}\right) \tag{4.3}\\
&  =\Omega_{i}\sum_{p=1}^{N}\frac{\gamma^{2}\left(  f_{i}\right)  }{\left(
f_{i}-\omega_{p}\right)  ^{2}}R_{p}\nonumber\\
R_{p}  &  =\frac{g^{2}}{2}\left(  \frac{\partial\omega_{p}}{\partial g}\right)
\nonumber
\end{align}

The $R_{p}$ satisfy the linear set of equations%

\begin{equation}
\left[  \sum_{k}\ \Omega_{k}\frac{\gamma^{2}\left(  f_{k}\right)  }{\left(
f_{k}-\omega_{p}\right)  ^{2}}+2s\sum_{q=1\left(  q\neq p\right)  }%
^{N^{\prime}}\frac{\gamma^{2}\left(  \omega_{q}\right)  }{\left(  \omega
_{p}-\omega_{q}\right)  ^{2}}\right]  R_{p}-2s\sum_{q=1\left(  q\neq p\right)
}^{N^{\prime}}\frac{\gamma^{2}\left(  \omega_{p}\right)  }{\left(  \omega
_{p}-\omega_{q}\right)  ^{2}}R_{q}=1 \tag{4.4}%
\end{equation}

obtained by differentiation of (3.11) with respect to $g$ .

\qquad The Hellman-Feynman theorem can also be used to calculate the
expectation values of the following operators%

\begin{align}
\frac{\partial K_{i}}{\partial f_{j}}  &  =g\frac{1}{\left(  f_{i}%
-f_{j}\right)  ^{2}}\frac{\gamma\left(  f_{i}\right)  }{\gamma\left(
f_{j}\right)  }K_{i,j}~,~i\neq j\tag{4.5}\\
K_{i,j}  &  =\left\{  u\left(  f_{i},f_{j}\right)  \left(  b_{i}^{\dag}%
~b_{j}+b_{j}^{\dag}~b_{i}\right)  -\frac{s}{2}\gamma\left(  f_{i}\right)
\gamma\left(  f_{j}\right)  \left(  \Omega_{i}+2~s~n_{i}\right)  \left(
\Omega_{j}+2~s~n_{j}\right)  \right\} \nonumber
\end{align}

or%

\begin{equation}
K_{i,j}=\frac{\left(  f_{i}-f_{j}\right)  ^{2}}{2g}\left[  \frac{\gamma\left(
f_{j}\right)  }{\gamma\left(  f_{i}\right)  }\frac{\partial K_{i}}{\partial
f_{j}}+\frac{\gamma\left(  f_{i}\right)  }{\gamma\left(  f_{j}\right)  }%
\frac{\partial K_{j}}{\partial f_{i}}\right]  \tag{4.6}%
\end{equation}

The expectation value of $\partial K_{i}/\partial f_{j}$ is given by%

\begin{equation}
\left\langle \psi\right\vert \frac{\partial K_{i}}{\partial f_{j}}\left\vert
\psi\right\rangle =\frac{\partial k_{i}}{\partial f_{j}} \tag{4.7}%
\end{equation}

and%

\begin{align}
\left\langle \psi\right\vert K_{i,j}\left\vert \psi\right\rangle  &
=\frac{\left(  f_{i}-f_{j}\right)  ^{2}}{2g}\left[  \frac{\gamma\left(
f_{j}\right)  }{\gamma\left(  f_{i}\right)  }\frac{\partial k_{i}}{\partial
f_{j}}+\frac{\gamma\left(  f_{i}\right)  }{\gamma\left(  f_{j}\right)  }%
\frac{\partial k_{j}}{\partial f_{i}}\right] \tag{4.8}\\
&  =-\Omega_{i}\Omega_{j}\gamma\left(  f_{i}\right)  \gamma\left(
f_{j}\right)  \left[  \frac{s}{2}+\frac{\left(  f_{i}-f_{j}\right)  ^{2}}%
{4}\sum_{p=1}^{N}\gamma^{2}\left(  \omega_{p}\right)  \left(  \frac{R_{p,j}%
}{\left(  f_{i}-\omega_{p}\right)  ^{2}}+\frac{R_{p,i}}{\left(  f_{j}%
-\omega_{p}\right)  ^{2}}\right)  \right] \nonumber
\end{align}

where we have written%

\begin{equation}
\frac{\partial\omega_{p}}{\partial f_{j}}=\Omega_{j}R_{p,j} \tag{4.9}%
\end{equation}

and where the $R_{p,j}$ satisfy the linear set of equations (4.4), with the
right-hand-side replaced by $\gamma^{2}\left(  \omega_{p}\right)  /\left(
f_{j}-\omega_{p}\right)  ^{2}$ , obtained by differentiating (3.11),

\bigskip

\section{Conclusions}

\qquad We have derived a general set of conditions for the integrability of
the most general seniority conserving Hamiltonian with two-body forces for
bosons or fermions. They have been used to obtain a new class of integrable
models. We have shown that the eigenvalues and eigenstates of the Hamiltonian
and the associated integrals of the motion can be obtained by the solution of
a fairly simple set of coupled algebraic equations. While this solvability
depends upon the integrability conditions in a rather complicated way, we have
not been able to construct a general result that integrability implies solvability.

The results of this paper can be generalized to other systems such as a system
of nucleons with an isospin-independent $J=0$ pairing in jj-coupling\cite{11}
or a spin- and isospin-independent $L=0$ pairing in ls-coupling.\cite{12}
Another system to which these methods can be applied is the system of fermions
with an equal mix of pairing in the $^{1}S_{0}$ and $^{3}P_{0}$
channels.\cite{13}


\begin{thebibliography}{99}                                                                                               %


\bibitem {1}J. Bardeen, L.N. Cooper, and J. R. Schriefer, Phys. Rev.
\textbf{106}, 162 (1957); Phys. Rev. \textbf{108,} 1175 (1957); R. P. Feynman,
\emph{Statistical Mechanics}, Ch. 10, W. A. Benjamin \& Co. (1972).

\bibitem {2}R. W. Richardson, Phys. Lett. \textbf{3}, 277 (1963); R. W.
Richardson and N. Sherman, Nucl. Phys. \textbf{52}, 221 (1964).

\bibitem {3}R. W. Richardson, J. Math. Phys. \textbf{6}, 1034 (1965).

\bibitem {4}R. W. Richardson, J. Math. Phys. \textbf{9}, 1327 (1968).

\bibitem {5}D. C. Ralph, C. T. Black, and M. Tinkham, Phys. Rev. Lett.
\textbf{74}, 3241 (1995); Phys. Rev. Lett. \textbf{76}, 688 (1996); Phys. Rev.
Lett. \textbf{78}, 4087 (1997).

\bibitem {6}J. von Delft and D. C. Ralph, Physics Reports \textbf{345}, 61 (2001).

\bibitem {7}M. C. Cambiaggio, A. M. F. Rivas, M. Saraceno, Nucl. Phys. A
\textbf{624}, 157 (1997).

\bibitem {8}L. Amico, A. Di Lorenzo, and A. Osterloh, Phys. Rev. Lett.
\textbf{86}, 5759 (2001); Nucl. Phys. B \textbf{614}, 499 (2001), e-print: cond-mat/0101029.

\bibitem {9}J. Dukelsky, C. Esebbag, and P. Schuck, Phys. Rev. Lett.
\textbf{87}, 66403 (2001), e-print: cond-mat/0107477.

\bibitem {10}R. W. Richardson, J. Math. Phys. \textbf{18}, 1802 (1977).

\bibitem {11}R. W. Richardson, Phys. Rev. \textbf{144}, 874 (1966).

\bibitem {12}R. W. Richardson, Phys. Rev. \textbf{159}, 792 (1967).

\bibitem {13}R. W. Richardson, Annals of Phys. \textbf{65}, 241 (1971).

\bibitem {14}C. Cohen-Tannoudji, B. Diu and F. Lalo, \emph{Quantum Mechanics},
John Wiley \& Sons, (1977), p. 1192 .
\end{thebibliography}
\end{document}